Title: Magnetic structure and double exchange in hypothetical compound $La_{1-x}Ca_xMnO_3$

Authors: B. V. Karpenko (1), L. D. Falkovskaya (1), and A. V. Kuznetsov (2)

( (1) Institute of Physics of Metals, Ural Branch of Russian Academy of Sciences, Ekaterinburg, Russia, (2) Ural State University, Ekaterinburg, Russia )

Comments: 17 pages with 14 figures

Subj-class: Strongly Correlated Electrons

Question about the appearance of different magnetic structures in the family of compounds $La_{1-x}Ca_xMnO_3$, $0 \leq x \leq 1$, is investigated theoretically. It is supposed that the whole series has $GdFeO_3$ structure type. The problem is considered at absolute zero temperature in the nearest neighbor approximation. Superexchange, double exchange and anisotropy energy are taken into account – all together 10 interaction parameters. The spin operator of double exchange interaction in crystal between ions with different valence $Mn^{3+}$ and $Mn^{4+}$ is the straight generalization of two-spin operator in the known problem of Anderson – Hasegawa molecule. Minimization of the ground state energy with respect to direction angles of magnetic sub-lattices leads to a system of transcendental equations whose solutions give 11 types of magnetic configurations : two ferromagnetic, three collinear antiferromagnetic and six non-collinear. When the concentration of Ca ions x changes one spin configuration replaces another as the ground state. As a whole the sequence of configurations when x changes from 0 till 1 can be brought in correspondence to those observed on the experiment. The comparison with experiment was made by means of "mixed" procedure: part of numerical values of interaction parameters and transition concentrations x from one configuration to another were taken from experiment while the missing values of these parameters were calculated with the help of the corresponding theoretical relations representing the conditions of equality of different phases energies at the concentration crossing.

PACS number(s): 75.47.Lx, 75.30.Et, 75.30.Kz

# I. INTRODUCTION

The perovskite-like manganese compounds were began to study as long ago as 1950 years.[1-3] A large amount of crystallographic and magnetic structures and variety of electric properties were revealed in $La_{1-x}Ca_xMnO_3$ and similar compounds. The correlation was found between the type of magnetic ordering and conducting properties of the substance. Apparently the manifestation of effect of the so called double exchange interaction predicted by Ziner[4,5] took place first in perovskites. Detail quantum-mechanical consideration of double-exchange mechanism on the example of molecule with two magnetic centers was made by Anderson and Hasegawa.[6] De Gennes[7] extended their results on the crystal.

Today the interest to perovskite-like systems appeared again due to the discovery of colossal magneto-resistance. The interest to the mechanism of double exchange and its role in magnetic, electric and magneto-resistant phenomena appeared again. In the present paper we are investigating the role of double exchange in forming of different magnetic phases in $La_{1-x}Ca_xMnO_3$.

As it was first shown in Ref.6 the main features of double exchange are the following. In the limit of large Hund interaction the ground state of the system is ferromagnetic, the energy depends upon the value of transfer integral for additional electron, the values of localized spin and the total molecule spin, but does not depend upon the value of Hund's exchange integral. The energy depends upon the total spin not as it square but linearly. This non-Heisenberg character of double exchange leads to the fact that magnetic susceptibility at high temperatures does not obey Curie-Weiss law. Besides in the classical limit exchange interaction is proportional to the value $\cos(\alpha/2)$, not $\cos\alpha$ (as in Heisenberg case), where $\alpha$ - is the angle between localized spins of different sites.

De Gennes[7] applied the double-exchange theory in the classical limit to a crystal with layered antiferromagnetic structure and showed that if the transfer energy for conduction electrons is proportional to $\cos(\alpha/2)$, while the interaction of localized spins is proportional to $\cos\alpha$ and is negative for sites of neighboring layers the appearance of non-collinear antiferromagnetic structure is possible in the system. In other words co-existence of Heisenberg and non-Heisenberg types of interaction in the system leads to the appearance of angular magnetic ordering.

In the present paper we are investigating the possibility of existence of different magnetic structures in $La_{1-x}Ca_xMnO_3$ with regard for nearest neighbors interactions of both Heisenberg and non- Heisenberg types. The consideration is made in quantum regime. The double - exchange interaction is described by an operator containing spin operators of different sites and having the square route character with respect to the product of spin operators. The double-exchange Hamiltonian for a crystal is a direct extension of double-exchange spin Hamiltonian for Anderson-Hasegawa molecule which is suitable for description of the lowest excited states. Different approximations used in our problem we shall discuss later.

## I I. ANDERSON-HASEGAWA MOLECULE AND EFFECTIVE SPIN HAMILTONIAN

Now we shall briefly set forth results for Anderson-Hasegawa molecule.[6] Let we have two centers *a* and *b* with localized on them spins $S_a = S_b = S$ and one additional electron (spin $\sigma = 1/2$ ), which can transfer from one center to another and has transfer integral $B > 0$. On each site spin $\vec{\sigma}$ is bound with spins $\vec{S}_a$ or $\vec{S}_b$ into rigid spin $S + 1/2$ due to the Hund's rule (the exchange integral $J$). Then as it is shown in Ref.6, the lower group of levels at $J \to \infty$ is given by the expression

$$E = -\frac{1}{2}\frac{2S_0 + 1}{2S + 1}B, \qquad (1)$$

where $S_0$ is the total molecule spin;

$$\vec{S}_0 = \vec{S}_a + \vec{S}_b + \vec{\sigma} ; \qquad (2)$$

$$S_0 = 1/2, 3/2, ..2S + 1/2, \qquad (3)$$

$2S+1$ values in all each of which is realized twice.

Let's use the identity

$$S_0 = -\frac{1}{2} + \sqrt{1+S(2S+3)}\sqrt{1+\frac{2}{1+S(2S+3)}(\vec{S}_a+\vec{\sigma})\vec{S}_b} \ . \tag{4}$$

Then energy (1) can be treated as the eigenvalues of an effective spin operator

$$\hat{H}_D = -pB\sqrt{1+q(\vec{S}_a+\vec{\sigma})\vec{S}_b} \ , \tag{5}$$

where

$$p = \frac{\sqrt{1+S(2S+3)}}{2S+1}, q = \frac{2}{1+S(2S+3)} . \tag{6}$$

(It is clear that Eqs. (4) and (5) can be presented in symmetrized form. )

The operator $\hat{H}_D$ from Eq.(5) we shall call the double exchange interaction operator. Its non-Heisenberg form is obvious. The operator $\hat{H}_D$ represents interaction between rigid spins $S+1/2$ and $S$. Just in this form we shall use it considering the crystal.

## III. HAMILTONIAN FOR THE CRYSTAL

Let's write down the model Hamiltonian for the crystal. We shall use the homogeneous model for magnetic manganese sub-lattice in $La_{1-x}Ca_xMnO_3$, that is we shall suppose that each site in the manganese sub-lattice is occupied with probability $x$ by ion $Mn^{4+}$ and has spin $S(Mn^{4+})=3/2$ and with probability $1-x$ - by ion $Mn^{3+}$ and has spin $S(Mn^{3+})=2$. Thus our magnetic system represents a homogeneous dynamical spin alloy of rigid spins $S_1=2$ and $S_2=3/2$. Further the nearest neighbors approximation will be used when each spin interacts with its nearest environment by means of two mechanisms: Heisenberg superexchange and non-Heisenberg double exchange.

Let's assume that the compound $La_{1-x}Ca_xMnO_3$ has the $GdFeO_3$ type of structure in the whole range $0 \leq x \leq 1$ (space group Pnma). The elementary cell of orthorhombic lattice is presented on Fig.1 where only manganese ions sites are presented. Numbers 1,2,3,4 enumerate four Bravais lattices. The base vectors of Bravais lattices (sub-lattices) are

$$\vec{\rho}_1 = 0, \vec{\rho}_2 = \frac{1}{2}(\vec{a}+\vec{b}), \vec{\rho}_3 = \frac{1}{2}\vec{c}, \vec{\rho}_4 = \frac{1}{2}(\vec{a}+\vec{b}+\vec{c}), \qquad (7)$$

where $\vec{a}, \vec{b}$ and $\vec{c}$ are vectors of primitive translations for a simple orthorhombic lattice. Vectors of six nearest neighbors are given by the expressions

$$\vec{\Delta}_1 = \frac{1}{2}(\vec{a}+\vec{b}), \vec{\Delta}_2 = \frac{1}{2}(-\vec{a}+\vec{b}), \vec{\Delta}_3 = -\frac{1}{2}(\vec{a}+\vec{b}), \vec{\Delta}_4 = \frac{1}{2}(\vec{a}-\vec{b}), \vec{\Delta}_5 = \frac{1}{2}\vec{c}, \vec{\Delta}_6 = -\frac{1}{2}\vec{c}. \qquad (8)$$

Let's denote the integral of superexchange interaction between ions $Mn^{3+} - Mn^{3+}$ in $(ab)$-plane as $I_1$ (the nearest neighbors $\vec{\Delta}_1, \vec{\Delta}_2, \vec{\Delta}_3, \vec{\Delta}_4$), while that along $c$-axis as $I_2$ (the nearest neighbors $\vec{\Delta}_5$ and $\vec{\Delta}_6$). Similarly for interaction $Mn^{4+} - Mn^{4+}$ in $(ab)$-plane as $I_3$, while along $c$-axis as $I_4$. Interactions $Mn^{3+} - Mn^{4+}$ in $(ab)$-plane we shall denote as $I_5$, and along $c$-axis as $I_6$. The transfer integral between ions $Mn^{3+}$ and $Mn^{4+}$ in $(ab)$-plane let's be $B_1$, and along $c$ - $B_2$. Besides we shall take into consideration anisotropy energy choosing $b$-axis as an easy one. We shall denote the anisotropy constant for ions $Mn^{3+}$ as $K_1 > 0$, and for ions $Mn^{4+}$ as $K_2 > 0$. Thus 10 interaction parameters are taken into account in our problem: $I_1, I_2, I_3, I_4, I_5, I_6, B_1, B_2, K_1, K_2$.

With regard for all stated above let's write down our model spin Hamiltonian $\hat{H}$ in the form:

$$\hat{H} = \hat{H}_{1ex} + \hat{H}_{2ex} + \hat{H}_a, \qquad (9)$$

where

$$\hat{H}_{1ex} = -\sum_{\vec{m}} \sum_{k=1}^{4} \sum_{i=1}^{6} \sum_{n,l=1}^{2} I_{nl}(\vec{\Delta}_i) \vec{S}_n(\vec{m}+\vec{\rho}_k) \vec{S}_l(\vec{m}+\vec{\rho}_k+\vec{\Delta}_i), \qquad (10)$$

$$\hat{H}_{2ex} = -\sum_{\vec{m}} \sum_{k=1}^{4} \sum_{i=1}^{6} \sum_{n,l=1}^{2} B_{nl}(\vec{\Delta}_i) \sqrt{1 + \frac{2}{1+S(2S+3)} \vec{S}_n(\vec{m}+\vec{\rho}_k) \vec{S}_l(\vec{m}+\vec{\rho}_k+\vec{\Delta}_i)}, \qquad (11)$$

$$\hat{H}_a = -\sum_{\vec{m}} \sum_{k=1}^{4} \sum_{n=1}^{2} A_n (S_n^z(\vec{m}+\vec{\rho}_k))^2, \qquad (12)$$

where $\hat{H}_{1ex}$ is the operator of superexchange interaction, $\hat{H}_{2ex}$ is the operator of double exchange, $\hat{H}_a$ is the operator of anisotropy energy. Summing over $\vec{m}$ in Eqs.

Eqs. (10-12) is led upon $N$ sites of sub-lattice (the whole number of sites is $4N$). Sum over index $k$ means summing over four sub-lattices. Index $i$ numbers the nearest neighbors. Indexes $n$ and $l$ distinguish ions $Mn^{3+}$ ($n,l = 1$) and $Mn^{4+}$ ($n,l = 2$). In this case $S_1 = S + (1/2)$ (spin of ion $Mn^{3+}$) and $S_2 = S$ (spin of ion $Mn^{4+}$, $S = 3/2$). The following notations are introduced also in Eqs. (10-12):

$$I_{11}(\vec{\Delta}_i) = (1-x)^2 I_1; i = 1,2,3,4. \tag{13}$$

$$I_{11}(\vec{\Delta}_i) = (1-x)^2 I_2; i = 5,6. \tag{14}$$

$$I_{22}(\vec{\Delta}_i) = x^2 I_3; i = 1,2,3,4. \tag{15}$$

$$I_{22}(\vec{\Delta}_i) = x^2 I_4; i = 5,6. \tag{16}$$

$$I_{12}(\vec{\Delta}_i) = I_{21}(\vec{\Delta}_i) = x(1-x)I_5; i = 1,2,3,4. \tag{17}$$

$$I_{12}(\vec{\Delta}_i) = I_{21}(\vec{\Delta}_i) = x(1-x)I_6; i = 5,6. \tag{18}$$

$$B_{11}(\vec{\Delta}_i) = B_{22}(\vec{\Delta}_i) = 0. \tag{19}$$

$$B_{12}(\vec{\Delta}_i) = B_{21}(\vec{\Delta}_i) = x(1-x)\sqrt{\frac{S+1}{2S+1}} B_1; i = 1,2,3,4. \tag{20}$$

$$B_{12}(\vec{\Delta}_i) = B_{21}(\vec{\Delta}_i) = x(1-x)\sqrt{\frac{S+1}{2S+1}} B_2; i = 5,6. \tag{21}$$

$$A_1 = (1-x)K_1, A_2 = xK_2. \tag{22}$$

The double exchange operator $\widehat{H}_{2ex}$ from Eq. (11) in the limit case of two sites goes to operator of Eq.(5) of Anderson-Hasegawa molecule.

For the following analysis it is convenient to introduce local coordinate systems according to Euler formulas for direction cosines $\alpha_{ij}$:

$\alpha_{11} = -\sin\phi \cos\chi - \cos\theta \cos\phi \sin\chi$,

$\alpha_{12} = \cos\phi \cos\chi - \cos\theta \sin\phi \sin\chi$,

$\alpha_{13} = \sin\theta \sin\chi$,

$\alpha_{21} = \sin\phi \sin\chi - \cos\theta \cos\phi \cos\chi$,

$\alpha_{22} = -\cos\phi \sin\chi - \cos\theta \sin\phi \cos\chi$, (23)

$\alpha_{23} = \sin\theta \cos\chi$,

$\alpha_{31} = \sin\theta \cos\phi$,

$$\alpha_{32} = \sin\theta \sin\phi,$$

$$\alpha_{33} = \cos\theta.$$

Later on we shall not investigate dependence upon angle $\chi$, leaving only polar and azimuth angles $\theta$ and $\phi$ and assuming $\chi = 0$. So, we have

$$S_n^{x_i}(\vec{m}+\vec{\rho}_k) = \sum_{j=1}^{3} \alpha_{ji}(k) S_n^{\xi_j(\vec{\rho}_k)}(\vec{m}+\vec{\rho}_k), x_1 \equiv x, x_2 \equiv y, x_3 \equiv z. \qquad (24)$$

Substituting Eq.(24) into Eqs.(10-12), we obtain

$$\widehat{H}_{1ex} = -\sum_{\vec{m}} \sum_{k=1}^{4} \sum_{i=1}^{6} \sum_{n,l=1}^{2} I_{nl}(\vec{\Delta}_i) \sum_{\delta,\lambda=1}^{3} \beta_{\delta\lambda}(\vec{\rho}_k, \vec{\Delta}_i) S_n^{\xi_\delta(\vec{\rho}_k)}(\vec{m}+\vec{\rho}_k) S_l^{\xi_\lambda(\vec{\rho}_k+\vec{\Delta}_i)}(\vec{m}+\vec{\rho}_k+\vec{\Delta}_i), \qquad (25)$$

$$\widehat{H}_{2ex} = -\sum_{\vec{m}} \sum_{k=1}^{4} \sum_{i=1}^{6} \sum_{n,l=1}^{2} B_{nl}(\vec{\Delta}_i) \times$$

$$\times \sqrt{1 + \frac{2}{(S+1)(2S+1)} \sum_{\delta,\lambda=1}^{3} \beta_{\delta\lambda}(\vec{\rho}_k, \vec{\Delta}_i) S_n^{\xi_\delta(\vec{\rho})}(\vec{m}+\vec{\rho}_k) S_l^{\xi_\lambda(\vec{\rho}_k+\vec{\Delta}_i)}(\vec{m}+\vec{\rho}_k+\vec{\Delta}_i)}, \qquad (26)$$

$$\widehat{H}_a = -\sum_{\vec{m}} \sum_{k=1}^{4} \sum_{n=1}^{2} A_n [\sum_{\delta=1}^{3} \alpha_{\delta 3}(\vec{\rho}_k) S_n^{\xi_\delta(\vec{\rho}_k)}(\vec{m}+\vec{\rho}_k)]^2, \qquad (27)$$

where

$$\beta_{\delta\lambda}(\vec{\rho}_k, \vec{\Delta}_i) = \sum_{r=1}^{3} \alpha_{\delta r}(\vec{\rho}_k) \alpha_{\lambda r}(\vec{\rho}_k+\vec{\Delta}_i). \qquad (28)$$

Eqs. (9, 25-28) present the final form of our Hamiltonian.

## IV. GROUND STATES ENERGIES OF DIFFERENT MAGNETIC CONFIGURATIONS

In the present paper we shall confine ourselves only to the case of absolute zero temperature. In order to consider energies of ground states we shall use the formula

$$S_n^{\xi_j(\vec{\rho}_k)}(\vec{m}+\vec{\rho}_k) = \delta_{j3} S_n, n = 1,2. \qquad (29)$$

Substituting Eq.(29) into Eqs.(9, 25-27), we obtain for the ground state energy $E$ the expression

$$E = -N\{a[\beta(12)+\beta(34)] + b[\beta(13)+\beta(24)] + c[\alpha^2(1)+\alpha^2(2)+\alpha^2(3)+\alpha^2(4)] +$$

$$+ 8Q[2(\sqrt{1+\gamma\beta(12)} + \sqrt{1+\gamma\beta(34)})B_1 + (\sqrt{1+\gamma\beta(13)} + \sqrt{1+\gamma\beta(24)})B_2]\}, \quad (30)$$

where

$$a = 8[(1-x)^2 S_1^2 I_1 + x^2 S_2^2 I_3 + 2x(1-x)S_1 S_2 I_5], \quad (31)$$

$$b = 4[(1-x)^2 S_1^2 I_2 + x^2 S_2^2 I_4 + 2x(1-x)S_1 S_2 I_6], \quad (32)$$

$$c = (1-x)S_1^2 K_1 + xS_2^2 K_2, \quad (33)$$

$$Q = x(1-x)\sqrt{\frac{S+1}{2S+1}}, \gamma = \frac{S}{S+1}, \quad (34)$$

$$\alpha(i) = \cos\theta_i, \beta(ij) = \sin\theta_i \sin\theta_j \cos(\phi_i - \phi_j) + \cos\theta_i \cos\theta_j, i,j = 1,2,3,4. \quad (35)$$

Indexes $i, j$ number sub-lattices.

The energy of Eq. (30) is minimal at angles $\theta$ and $\phi$, satisfying eight equations:

$$\frac{\partial E}{\partial \theta_1} = \frac{\partial E}{\partial \theta_2} = \frac{\partial E}{\partial \theta_3} = \frac{\partial E}{\partial \theta_4} = \frac{\partial E}{\partial \phi_1} = \frac{\partial E}{\partial \phi_2} = \frac{\partial E}{\partial \phi_3} = \frac{\partial E}{\partial \phi_4} = 0. \quad (36)$$

The system (36) has the following solutions.

$$A: \phi_1 = \phi_2 = \phi_3 = \phi_4 = \phi; \theta_1 = \theta_2 = 0, \theta_3 = \theta_4 = \pi. \quad (37)$$

$$A_1: \phi_1 = \phi_2 = \phi_3 = \phi_4 = \phi, \theta_1 = \theta_2 = \frac{1}{2}\arccos h, \theta_3 = \theta_4 = \pi - \theta_1. \quad (38)$$

$$A_2: \phi_1 = \phi_2 = \phi, \phi_3 = \phi_4 = \phi \pm \pi, \theta_1 = \theta_2 = \theta_3 = \theta_4 = \frac{1}{2}\arccos(-h_1). \quad (39)$$

$$B: \phi_1 = \phi_2 = \phi_3 = \phi_4 = \phi, \theta_1 = \theta_2 = \theta_3 = \theta_4 = 0. \quad (40)$$

$$B': \phi_1 = \phi_2 = \phi_3 = \phi_4 = \phi, \theta_1 = \theta_2 = \theta_3 = \theta_4 = \pi/2. \quad (41)$$

$$C: \phi_1 = \phi_2 = \phi_3 = \phi_4 = \phi, \theta_1 = \theta_3 = 0, \theta_2 = \theta_4 = \pi. \quad (42)$$

$$C_1: \phi_1 = \phi_2 = \phi_3 = \phi_4 = \phi, \theta_1 = \theta_3 = \frac{1}{2}\arccos g, \theta_2 = \theta_4 = \pi - \theta_1. \quad (43)$$

$$C_2: \phi_1 = \phi_3 = \phi, \phi_2 = \phi_4 = \phi \pm \pi, \theta_1 = \theta_2 = \theta_3 = \theta_4 = \frac{1}{2}\arccos(-g_1). \quad (44)$$

$$G: \phi_1 = \phi_2 = \phi_3 = \phi_4 = \phi, \theta_1 = \theta_4 = 0, \theta_2 = \theta_3 = \pi. \quad (45)$$

$$G_1: \phi_1 = \phi_2 = \phi_3 = \phi_4 = \phi, \theta_1 = \theta_4 = \frac{1}{2}\arccos h_2, \theta_2 = \theta_4 = \pi - \theta_1. \quad (46)$$

$$G_2: \phi_1 = \phi_4 = \phi, \phi_2 = \phi_3 = \phi \pm \pi, \theta_1 = \theta_2 = \theta_3 = \theta_4 = \frac{1}{2}\arccos(-g_2). \quad (47)$$

(Angle $\phi$ is arbitrary. Here and below symbols *A*, *B* and so on introduced by Wollan and Koehler in Ref.3 are used in order to denote spin configurations.) The following notations are input above:

$$h = \frac{1}{\gamma}\left[1 - \left(\frac{f}{b-c}\right)^2\right], \qquad (48)$$

$$h_1 = \frac{1}{\gamma}\left[1 - \left(\frac{f}{b+c}\right)^2\right], \qquad (49)$$

$$g = \frac{1}{\gamma}\left[1 - \left(\frac{d}{a-c}\right)^2\right], \qquad (50)$$

$$g_1 = \frac{1}{\gamma}\left[1 - \left(\frac{d}{a+c}\right)^2\right], \qquad (51)$$

$$h_2 = \frac{1}{\gamma}\left[1 - \left(\frac{d+f}{a+b-c}\right)^2\right], \qquad (52)$$

$$g_2 = \frac{1}{\gamma}\left[1 - \left(\frac{d+f}{a+b+c}\right)^2\right], \qquad (53)$$

where

$$d = \frac{8Sx(1-x)B_1}{\sqrt{(S+1)(2S+1)}}, \qquad (54)$$

$$f = \frac{4Sx(1-x)B_2}{\sqrt{(S+1)(2S+1)}}. \qquad (55)$$

We have for energies of the corresponding states $\varepsilon \equiv E/4N$

$$\varepsilon_A = -\frac{1}{2}\left[a - b + 2c + 8\left(2B_1 + \frac{B_2}{\sqrt{2S+1}}\right)x(1-x)\right], \qquad (56)$$

$$\varepsilon_{A_1} = -\frac{1}{2}\left[a - \frac{S+1}{S}b + \frac{2S+1}{S}c + 16B_1 x(1-x) - \frac{16B_2^2 S}{(2S+1)(b-c)}x^2(1-x)^2\right], b-c < 0, \qquad (57)$$

$$\varepsilon_{A_2} = -\frac{1}{2}\left[a - \frac{S+1}{S}b - \frac{1}{S}c + 16B_1 x(1-x) - \frac{16SB_2^2}{(2S+1)(b+c)}x^2(1-x)^2\right], b+c < 0, \qquad (58)$$

$$\varepsilon_B = -\frac{1}{2}[a + b + 2c + 8(2B_1 + B_2)x(1-x)], \qquad (59)$$

$$\varepsilon_{B'} = -\frac{1}{2}[a + b + 8(2B_1 + B_2)x(1-x)], \qquad (60)$$

$$\varepsilon_C = -\frac{1}{2}\left[-a+b+2c+8\left(\frac{2B_1}{\sqrt{2S+1}}+B_2\right)x(1-x)\right], \tag{61}$$

$$\varepsilon_{C_1} = -\frac{1}{2}\left[-\frac{S+1}{S}a+b+\frac{2S+1}{S}c-\frac{64SB_1^2}{(2S+1)(a-c)}x^2(1-x)^2+8B_2x(1-x)\right], a-c<0, \tag{62}$$

$$\varepsilon_{C_2} = -\frac{1}{2}\left[-\frac{S+1}{S}a+b-\frac{1}{S}c-\frac{64SB_1^2}{(2S+1)(a+c)}x^2(1-x)^2+8B_2x(1-x)\right], a+c<0, \tag{63}$$

$$\varepsilon_G = -\frac{1}{2}\left[-a-b+2c+\frac{8}{\sqrt{2S+1}}(2B_1+B_2)x(1-x)\right], \tag{64}$$

$$\varepsilon_{G_1} = -\frac{1}{2}\left[-\frac{S+1}{S}a-\frac{S+1}{S}b+\frac{2S+1}{S}c-\frac{16S(2B_1+B_2)^2}{(2S+1)(a+b-c)}x^2(1-x)^2\right], a+b-c<0, \tag{65}$$

$$\varepsilon_{G_2} = -\frac{1}{2}\left[-\frac{S+1}{S}a-\frac{S+1}{S}b-\frac{1}{S}c-\frac{16S(2B_1+B_2)^2}{(2S+1)(a+b+c)}x^2(1-x)^2\right], a+b+c<0. \tag{66}$$

Thus we have obtained five collinear magnetic configurations two of which ($B$ and $B'$) are ferromagnetic while three are antiferromagnetic ($A$, $C$, $G$), and also six non-collinear structures ($A_1, A_2, C_1, C_2, G_1, G_2$). In the ferromagnetic phase $B$ vector of magnetization is directed along the easy axis $b$, while in ferromagnetic phase $B'$ vector of magnetization is perpendicular to axis $b$. The collinear phases $A, B, B', C, G$ are depicted in Fig. 2. In $A$ state each spin is surrounded by four parallel to it spins and two antiparallel. In $C$ state the surrounding consists of two parallel and four antiparallel spins. In configuration $G$ all six nearest neighboring spins are antiparallel to the central spin. The figure shows that in $B'$ state spins are parallel to $c$-axis, however this is not necessarily – they must only lie in the $(ac)$-plane. In non-collinear phases $A_1$, $C_1$, $G_1$ vectors of spin sub-lattices are situated symmetrically with respect to the $b$-axis normal, while in configurations $A_2$, $C_2$, $G_2$ the sub-lattices spins are situated symmetrically with respect to $b$-axis as it is shown in Fig.3. On this Figure $\Theta$ is the angle of non-collinearity (the angle of canting), and $\Phi$ is the angle between sub-lattice vectors.

Energies of Eqs. (56-66) as the functions of concentration $x$ are shown in Figs.4, 5, 6, 7 at certain values of interaction parameters (see the next section).

## V. COMPARISON WITH EXPERIMENT

Numerous experiments demonstrate that with increase of $x$ in $La_{1-x}Ca_xMnO_3$ magnetic configurations change each other. Being guided by the results of Refs. 1, 2, 3, 8-16, we can suppose the following scheme of magnetic transitions when $x$ changes from 0 till 1:

$$A \to A_1 \to A_2 \to B \to C_2 \to C_1 \to C \to G. \tag{67}$$

The following equations should be held in the transition points

$$\varepsilon_A(x_1) = \varepsilon_{A_1}(x_1), \tag{68}$$

$$\varepsilon_{A_1}(x_2) = \varepsilon_{A_2}(x_2), \tag{69}$$

$$\varepsilon_{A_2}(x_3) = \varepsilon_B(x_3), \tag{70}$$

$$\varepsilon_B(x_4) = \varepsilon_{C_2}(x_4), \tag{71}$$

$$\varepsilon_{C_2}(x_5) = \varepsilon_{C_1}(x_5), \tag{72}$$

$$\varepsilon_{C_1}(x_6) = \varepsilon_C(x_6), \tag{73}$$

$$\varepsilon_C(x_7) = \varepsilon_G(x_7). \tag{74}$$

Substituting energies of Eqs. (56-66) into Eqs. (68-74), we obtain

$$b(x_1) - c(x_1) + \frac{4B_2 S x_1 (1 - x_1)}{\sqrt{2S+1}} = 0, A \to A_1. \tag{75}$$

$$b^2(x_2) - c^2(x_2) - \frac{16 S^2 B_2^2 x_2^2 (1 - x_2)^2}{(S+1)(2S+1)} = 0, A_1 \to A_2. \tag{76}$$

$$b(x_3) + c(x_3) + \frac{4S B_2 x_3 (1 - x_3)}{2S+1} = 0, A_2 \to B. \tag{77}$$

$$a(x_4) + c(x_4) + \frac{8S B_1 x_4 (1 - x_4)}{2S+1} = 0, B \to C_2. \tag{78}$$

$$a^2(x_5) - c^2(x_5) - \frac{64 S^2 B_1^2 x_5^2 (1 - x_5)^2}{(S+1)(2S+1)} = 0, C_2 \to C_1. \tag{79}$$

$$a(x_6) - c(x_6) + \frac{8 B_1 S x_6 (1 - x_6)}{\sqrt{2S+1}} = 0, C_1 \to C. \tag{80}$$

$$b(x_7) + 4B_2 \left(1 - \frac{1}{\sqrt{2S+1}}\right) x_7 (1 - x_7) = 0, C \to G. \tag{81}$$

So we have seven Eqs. (75-81) with seventeen unknowns: ten parameters $I_1 - I_6$, $K_1$, $K_2$, $B_1$, $B_2$ and seven transition concentrations $x_1 - x_7$. In order to use the system of Eqs. (75-81) in practice we should determine some ten values independently, while the missing seven ones we shall find solving the system of seven Eqs. (75-81). Let's do it in the following way. The values $I_1 = 9.6K$, $I_2 = -6.7K$, $K_1 = 1.92K$ were obtained in Ref.8 during the analysis of spin-wave spectrum in LaMnO$_3$. We arbitrarily (not having experimental data) assumed $K_2 = K_1 = 1.92K$. From data on Neel temperature for CaMnO$_3$[3,16] we shall put approximately $I_3 = I_4 = -8.7K$. Keeping in mind the results of papers [3,12-16] let's put some mean experimental values for transition concentrations $x_3 = 0.2$, $x_4 = 0.5$, $x_6 = 0.7$, $x_7 = 0.85$. Substituting these ten parameters $I_1, I_2, I_3, I_4, K_1, K_2, x_3, x_4, x_6, x_7$ into the system of Eqs. (75-81), we have obtained for the missing parameters the following values:

$$I_5 = -12.673K, I_6 = 9.603K, B_1 = 144.571K, B_2 = 116.085K, \tag{82}$$

$$x_1 = 0.173, x_2 = 0.197, x_5 = 0.55. \tag{83}$$

Using these values we have depicted plots of energies in the whole range of $x$ values in Figs. 4-7. Figure 8 shows the continuous curve $\varepsilon(x)$, made up of the pieces of functions $\varepsilon_A$, $\varepsilon_{A_1}$, $\varepsilon_{A_2}$, $\varepsilon_B$, $\varepsilon_{C_2}$, $\varepsilon_{C_1}$, $\varepsilon_C$ and $\varepsilon_G$, closing each other consecutively in the transition points $x_1$, $x_2$, $x_3$, $x_4$, $x_5$, $x_6$ and $x_7$. Figure 9 shows the dependence $\theta_1(x)$. Figure 10 depicts the dependence of angle $\Phi$ between sub-lattices moments upon $x$, Fig.11 is the plot of canting angle $\Theta(x)$, while Fig.12 presents the concentration dependence of ferromagnetic momentum (on one ion of manganese) $m = 2\mu_B S_F$, where $\mu_B$ is Bohr magneton and $S_F$ is given by formula

$$S_F = \frac{1}{2}(4-x)\cos\frac{\Phi}{2}. \tag{84}$$

Functions $\varepsilon(x)$ for collinear states $A, B, C, G$ have positive second derivatives $\partial^2\varepsilon/\partial x^2$ in the whole range $0 \leq x \leq 1$. The energies of non-collinear states $\varepsilon_{A_1}, \varepsilon_{A_2}, \varepsilon_{C_1}, \varepsilon_{C_2}, \varepsilon_{G_1}, \varepsilon_{G_2}$ are defined each in the own region of definition

$$\varepsilon_{A_1}: 0.173 \leq x \leq 0.239, \tag{85}$$

$$\varepsilon_{A_2}: 0.147 \leq x \leq 0.2, \tag{86}$$

$$\varepsilon_{C_1}: \quad 0.474 \leq x \leq 0.7, \tag{87}$$

$$\varepsilon_{C_2}: \quad 0.5 \leq x \leq 0.721, \tag{88}$$

$$\varepsilon_{G_1}: \quad 0.594 \leq x \leq 0.786, \tag{89}$$

$$\varepsilon_{G_2}: \quad 0.617 \leq x \leq 0.797. \tag{90}$$

The states $A_1, A_2, G_1, G_2$ have negative second derivative $\partial^2\varepsilon/\partial x^2$ in their definition regions. The state $C_1$ has negative second derivative for $x < 0.515$ and positive one for $x > 0.515$. Respectively for $C_2$ the point of inflection for $\varepsilon(x)$ is equal to $0.54$. Thus in the range $x_5 \leq x \leq x_6$ for $C_1$ the second derivative is positive, while for $C_2$ the second derivative changes sign from negative to positive with the increase of $x$ in the range $x_4 \leq x \leq x_5$.

It was experimentally shown in Ref.12, that transition from antiferromagnetic phase $A$ at small $x$ with increase of concentration to ferromagnetic state with $\theta_1 = \Theta = 90^0$ (in our terminology $B'$) occurs rather smoothly (see Fig.13 which should be compared with our Fig.11) and only through the sequence of one-type angle states (analog of our configuration $A_1$). So we see that unfortunately the results of our theory and this experiment differ essentially: state $B'$ does not realize in our theory at all and two intermediate angle states $A_1$ and $A_2$ exist between antiferromagnetic state $A$ and ferromagnetic state $B$ ($\Theta = 0$). Besides the range of transition between anti- and ferromagnetic states in Ref.12 is much broader than we have obtained. We do not have enough experimental data about the dependence of canting angle upon concentration $\Theta(x)$ for values $x \geq 0.5$ in order to compare in details theory and experiment. Only Ref.13 informs about angle magnetism at $x = 0.667$.

Our theoretical curve for $m$ (see Fig.12) can be compared with experimental data from Refs. 1,3, depicted on Fig.14. The type of the dependence is approximately the same while the qualitative coincidence is not good enough.

## VI. DISCUSSION

The proposed model of homogeneous dynamical spin alloy for compound La$_{1-x}$Ca$_x$MnO$_3$ has in principle allowed to explain the different magnetic configurations observed experimentally and the sequence of their change depending on the concentration $x$. The numerical values of interaction parameters obtained theoretically are quite plausible. Taking account of simultaneous presence of Heisenberg-type superexchange and non-Heisenberg –type double exchange leads to the appearance of non-collinear (canted) structures. Superexchange interaction can be both positive (ferromagnetic) and negative (antiferromagnetic). Double exchange is essentially positive. Canted structures appear at values of concentration $x$, when balance takes place between the total antiferromagnetic superexchange interaction and ferromagnetic double exchange. Superexchange interaction and double exchange taken separately cannot lead to canted configurations.

Presence of anisotropy energy (which is small) leads to the fact that angle states "symmetrical" with respect to the normal of $b$-axis turn at certain $x$ values into states "symmetrical" with respect to $b$-axis. At finite values of anisotropy energies ferromagnetic state $B'$, which is the limit case of angle states ($\theta = 90^0$), symmetrical with respect to normal of $b$-axis turns out to be unattainable. Ferromagnetic state $B$ always has lower energy than that of ferromagnetic state $B'$ and their difference is just anisotropy energy. Let's note that the canting angle $\Theta(x)$ has infinite derivatives with respect to $x$ in the points of transitions between collinear and non-collinear structures ($x_1, x_3, x_4, x_6$).

Let's note that magnetic phase transitions depending upon concentration in systems with double exchange were investigated earlier by many authors with the use of band approximation which was proposed by de-Gennes in his famous work[7]. The typical feature of this approximation is the proposal that additional electron realizing exchange coupling between localized spins is the conduction electron whose transfer integral depends upon mutual orientation of localized electrons spins. The band description is used for it so it is natural to call such an approach to double exchange theory the band approximation of double exchange. Then the approximation used in our paper can be relevantly called the localized approximation of double exchange though the conditionality of such a title is obvious. We did not use the concept of bands and only different spins $S+1/2$ and $S$ for ions $Mn^{3+}$ and $Mn^{4+}$ are present in our approach. (However this does not mean that our model forbids electronic conductivity. The conductivity is "hiding" here in probabilities $x$ and $1-x$ which we have introduced. Of course they allow electron migration from site to site. It seems that in our approach it's more proper to speak about not band but jump conductivity.) The difference between ions $Mn^{3+}$ and $Mn^{4+}$ disappears in band model and correspondingly different superexchange parameters between ions $Mn^{3+}$ and $Mn^{4+}$ are absent while in our approach these parameters are present separately. This essential difference does not allow compare literally the band and localized approximations.

The following moments should be associated with difficulties of our theory. The negativity of the second derivative $\partial^2\varepsilon/\partial x^2$ for angle states $A_1$, $A_2$ and partly $C_2$ indicates non-stability of these states. This question needs an additional investigation. It's impossible to obtain phenomenon of charge ordering in our approach because we are using homogeneous approximation. In connection with this we have simplified the situation supposing that only configurations $C_1$, $C_2$ or C are present in the range $0.5 > x > 0.85$ while experiment shows that CE states are also present in this range of $x$. We have called the object under investigation "*hypothetical* compound $La_{1-x}Ca_xMnO_3$" because of the fact that really at $x$ increasing from 0 till 1 both the lattice symmetry change (LaMnO₃ has rhombic symmetry while CaMnO₃ is cubic) and the lattice parameters hence the exchange parameters $I$, $B$, $K$ also change. Unfortunately the dependence of interaction parameters upon $x$ is unknown. It's of course inconsistently to assume parameters $I_1$ and $I_2$ to be different while $I_3$ and $I_4$ equal ones as we have done above at numerical evaluations. Thus numerical values for $I$, $B$, $K$ obtained in chapter V are approximate and averaged .

# ACKNOWLEDGEMENTS

This work was supported by Federal Scientific Technical Programme of Ministry of Science N 40.012.1.1.1153 Department of Physical Science of Russian Academy of Sciences 19, 490 and by Programme of the Presidium of the RAS "New materials and structures".

The authors would like to thank L.V. Karpenko for her assistance in the article arranging.

# REFERENCES

[1] G. H. Jonker and J. H. van Santen, Physica **16**, 337 (1950).
[2] G. H. Jonker and J. H. van Santen, Physica **16**, 599 (1950).
[3] E. O. Wollan and W. C. Koehler, Phys. Rev. **100**, 545 (1955).
[4] C. Zener, Phys. Rev. **81**, 440 (1951).
[5] C. Zener, Phys. Rev. **82**, 403 (1951).
[6] P. W. Anderson and H. Hasegawa, Phys. Rev. **100**, 675 (1955).
[7] P.-G. de Gennes, Phys. Rev. **118**, 141 (1960).
[8] F. Moussa, M. Hennion, J. Rodriguez-Carvajal, H. Moudden, L. Pinsard and A. Revcolevschi, Phys. Rev. B **54**, 15149 (1996).
[9] M. Hennion, F. Moussa, J. Rodriguez-Carvajal, L. Pinsard and A. Revcolevschi, Phys. Rev. B **56**, R497 (1997).
[10] M. Hennion, F. Moussa, G. Bioteau, J. Rodriguez-Carvajal, L. Pinsard and A. Revcolevschi, Phys. Rev. Lett. **81**, 1957 (1998).
[11] M. Hennion, F. Moussa. G. Biotteau, J. Rodriguez-Carvajal, L. Pinsard and A. Revcolevschi, Phys. Rev. B **61**, 9513 (1999).
[12] G. Biotteau, M. Hennion, F. Moussa, J. Rodriguez-Carvajal, L. Pinsard, A. Revcolevschi, Y. M. Mukovskii and D. Shulyatev, Phys. Rev. B **64** 104421 (2001).
[13] P.G.Radaelli, D. E. Cox, L. Capogna, S.-W. Cheong and M. Marezio, Phys. Rev. B **59** 14440 (1999).
[14] M. Pissas, G. Kallias, M. Hofmann and D. M. Többens, Phys. Rev. B **65**, 064413 (2002).
[15] M. Pissas and G. Kallias, Phys. Rev. B **68**, 134414 (2003).
[16] R. H. Heffner, J. E. Sonier, D. E. Machlaughlin, G. J. Nieuwenhuys, G. M. Luke, Y. J. Uemura, William Ratcliff II, S. W, Cheong and G. Balakrishnan, Phys. Rev. B **63**, 094408 (2001).

# FIGURE CAPTIONS

Fig.1. The elementary cell of $La_{1-x}Ca_xMnO_3$. Only Mn-sites are denoted. Figures 1-4 number Bravais lattices.

Fig.2. Collinear magnetic configurations $A, B, B', C, G$. Arrows denote the site spins.

Fig.3. The ordering of sub-lattice moments in angle configurations $A_1, C_1, G_1$ and $A_2, C_2, G_2$.

Fig.4. Energies $\varepsilon(x)$ for collinear structures $A, B, B', C, G$.

Fig.5. Energies $\varepsilon(x)$ for non-collinear structures $A_1$ (thin line) and $A_2$ (thick line).

Fig.6. Energies $\varepsilon(x)$ or non-collinear structures $C_1$ (thin line) and $C_2$ (thick line).

Fig.7. Energies $\varepsilon(x)$ or non-collinear structures $G_1$ (thin line) and $G_2$ (thick line).

Fig.8. Energy $\varepsilon(x)$, consisting of energies $\varepsilon_A(x)$, $\varepsilon_{A_1}(x)$, $\varepsilon_{A_2}(x)$, $\varepsilon_B(x)$, $\varepsilon_{C_2}(x)$, $\varepsilon_{C_1}(x)$, $\varepsilon_C(x)$ and $\varepsilon_G(x)$. Vertical lines are drawn through transition points $x_1 - x_7$.

Fig.9. Dependence of angle $\theta_1$ upon concentration $x$.

Fig.10. Dependence of angle $\Phi$ between sub-lattices vectors upon concentration $x$.

Fig.11. Theoretical dependence of canting angle $\Theta$ upon concentration $x$.

Fig.12. Theoretical dependence of ferromagnetic momentum $m$ upon concentration $x$.

Fig.13. Experimental dependence of angle $\Theta$ upon concentration in $La_{1-x}Ca_xMnO_3$. The plot is taken from Ref.12.

Fig.14. Experimental dependence of ferromagnetic momentum $m$ per one Mn ion upon concentration $x$ for $La_{1-x}Ca_xMnO_3$. (See Fig.12 of Ref.3). Clear circles are data of Ref.1, while dark ones were taken from Ref.3.

**Fig. 1**

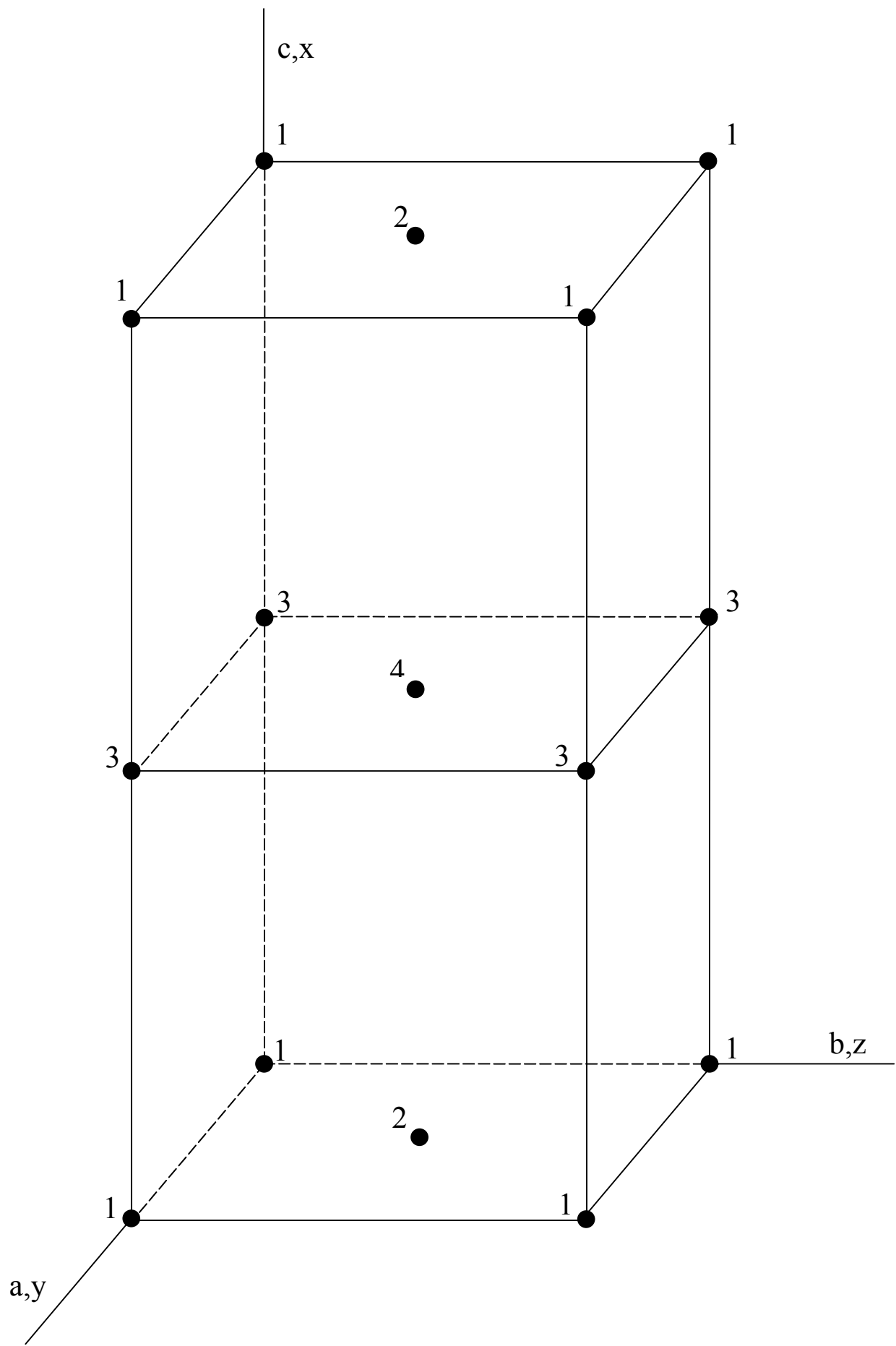

**Fig. 2**

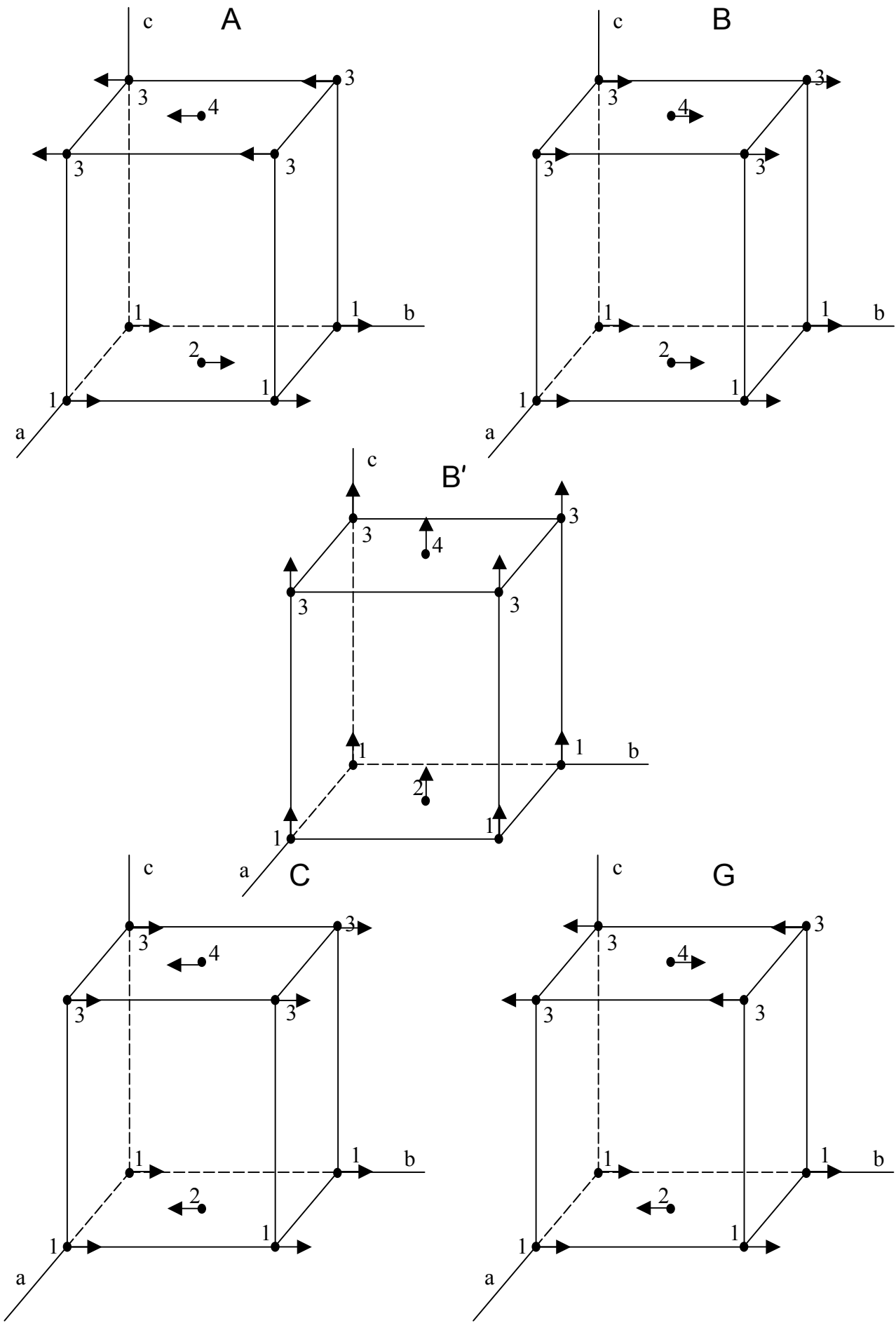

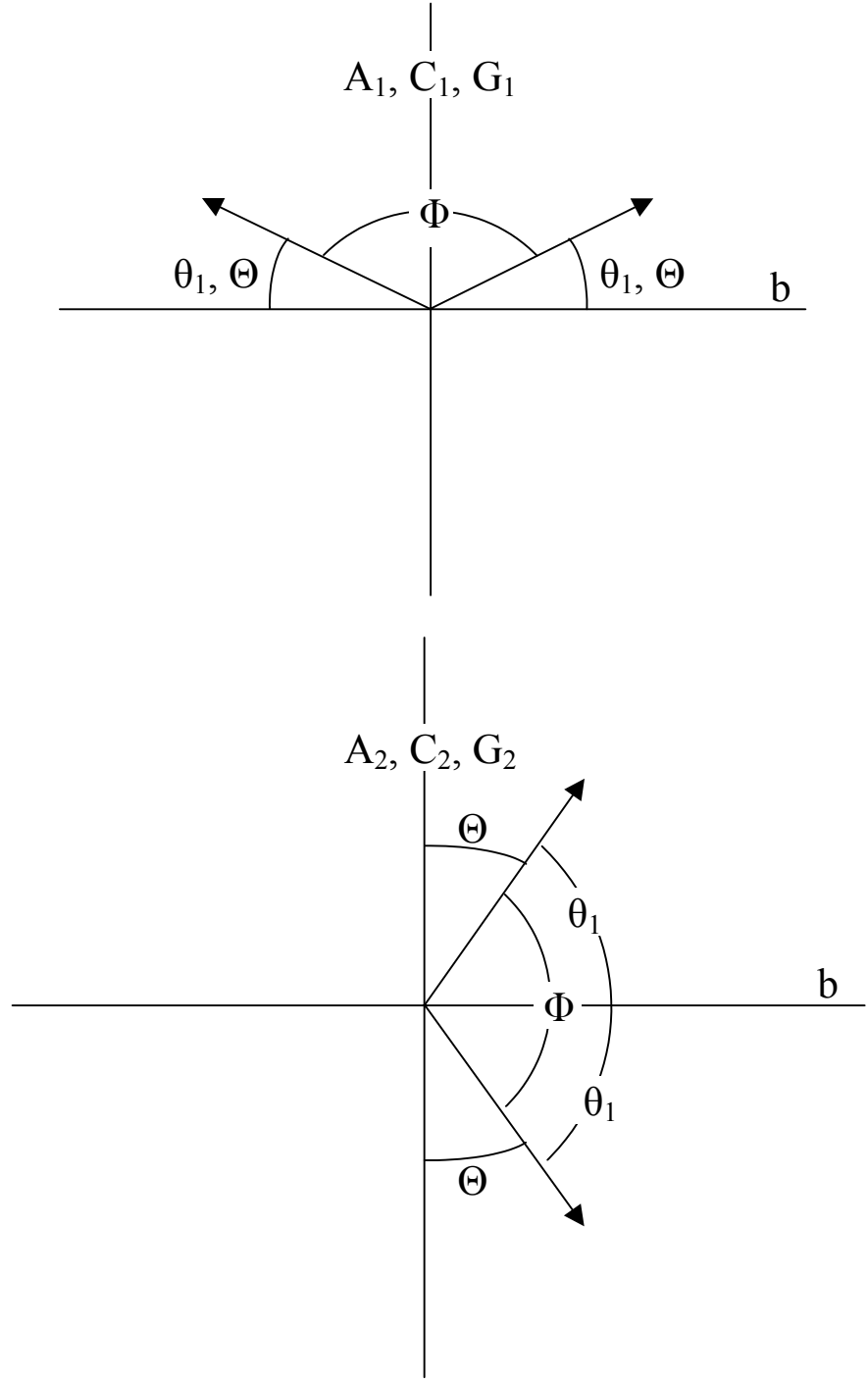

**Fig.3**

**Fig. 4**

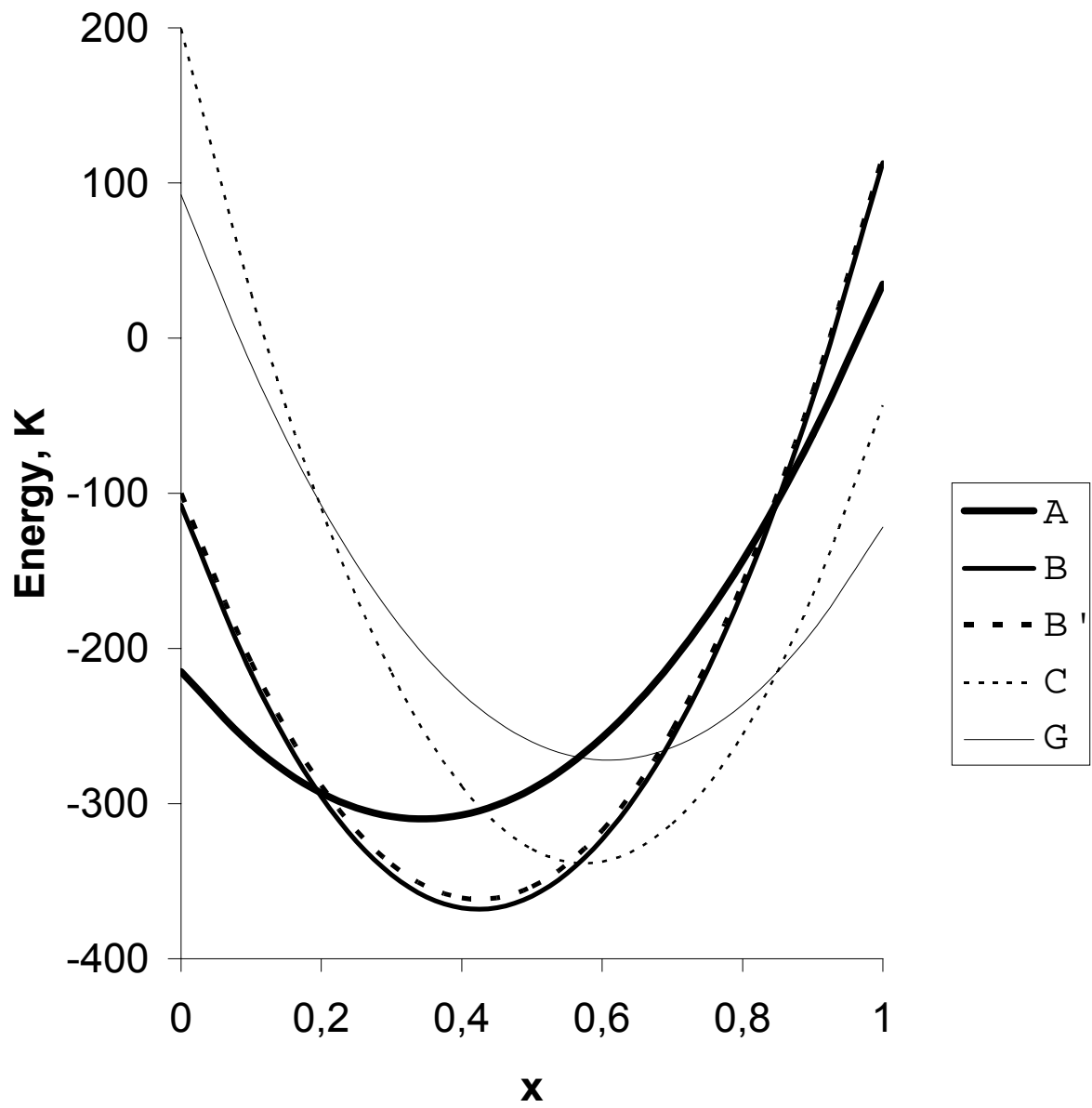

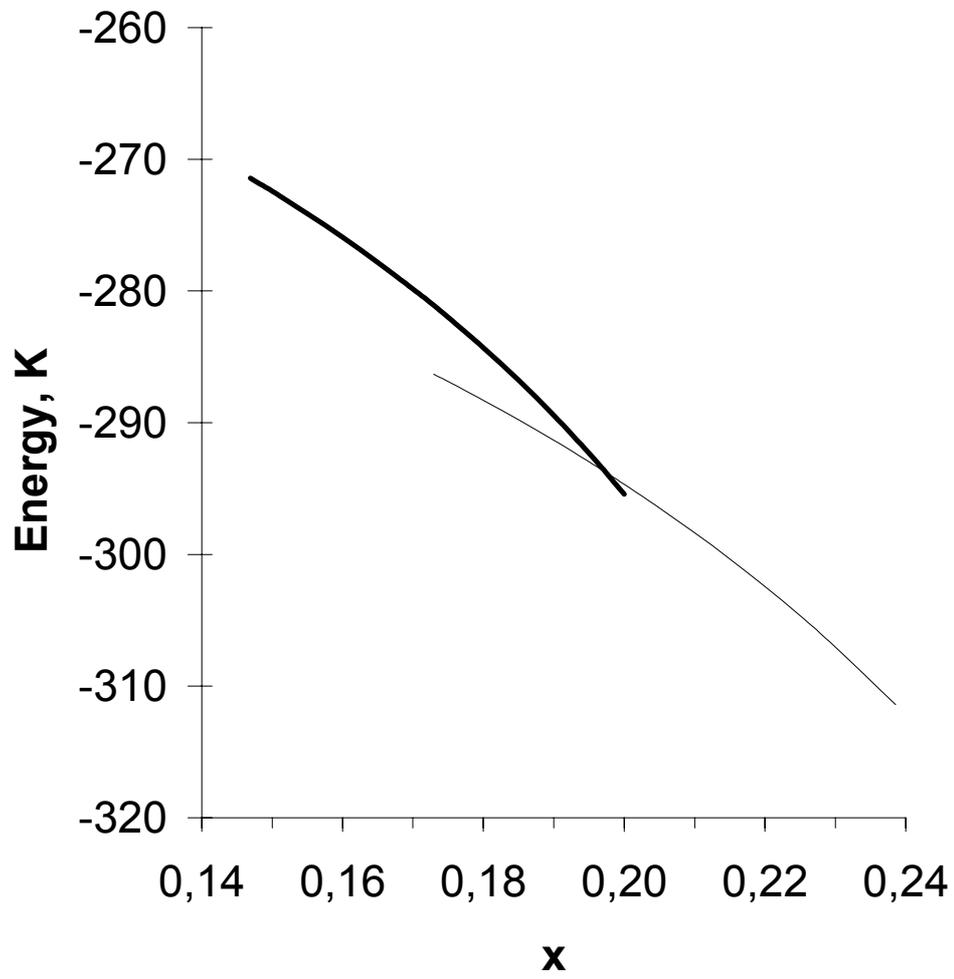

**Fig. 5**

**Fig. 6**

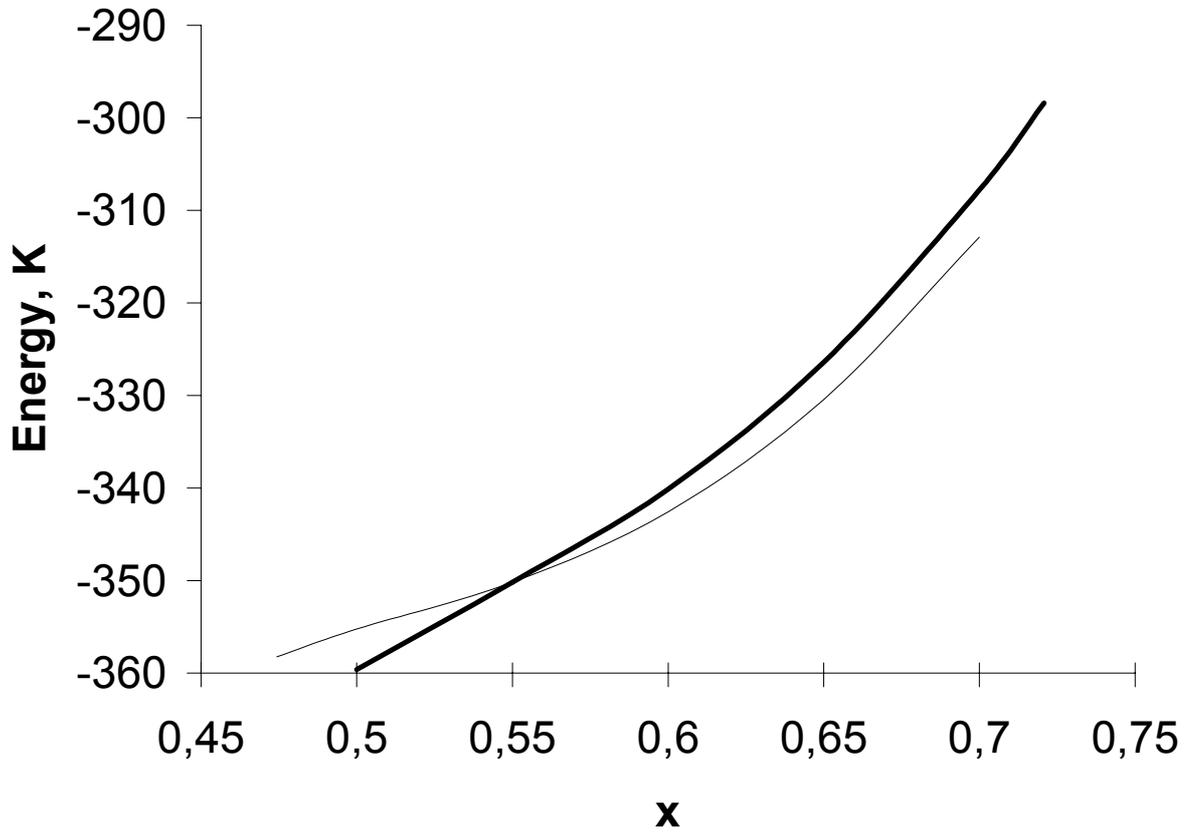

**Fig. 7**

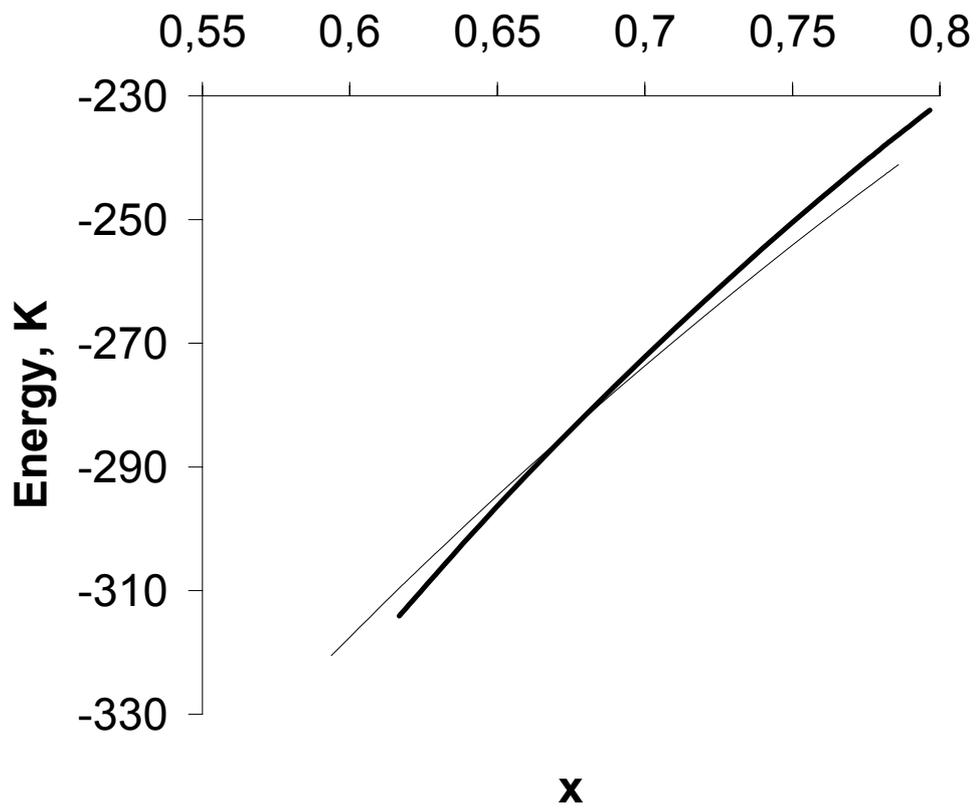

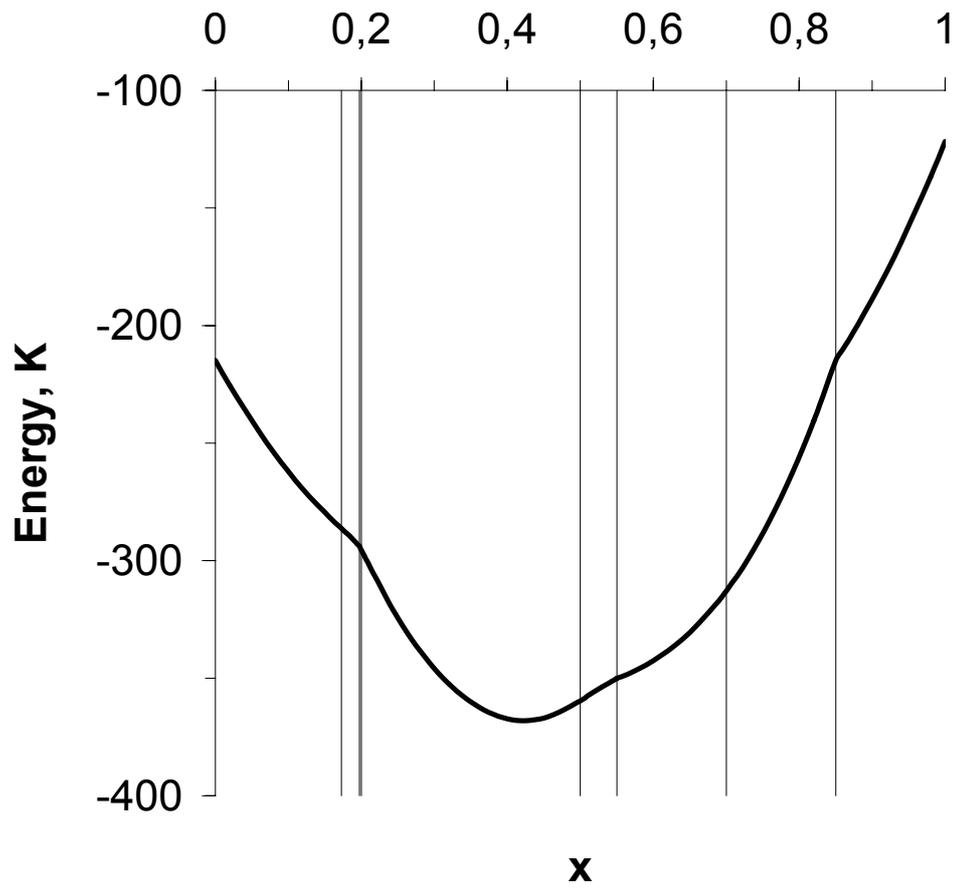

**Fig.8**

**Fig. 9**

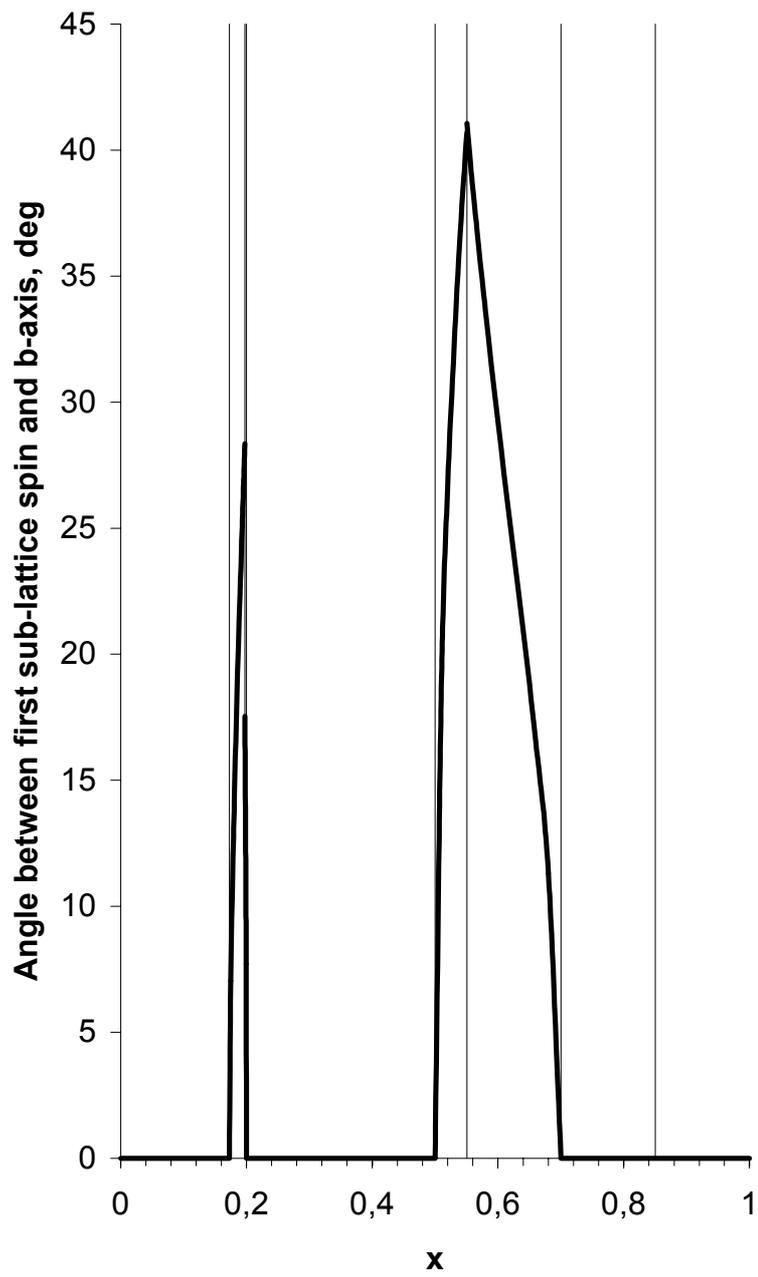

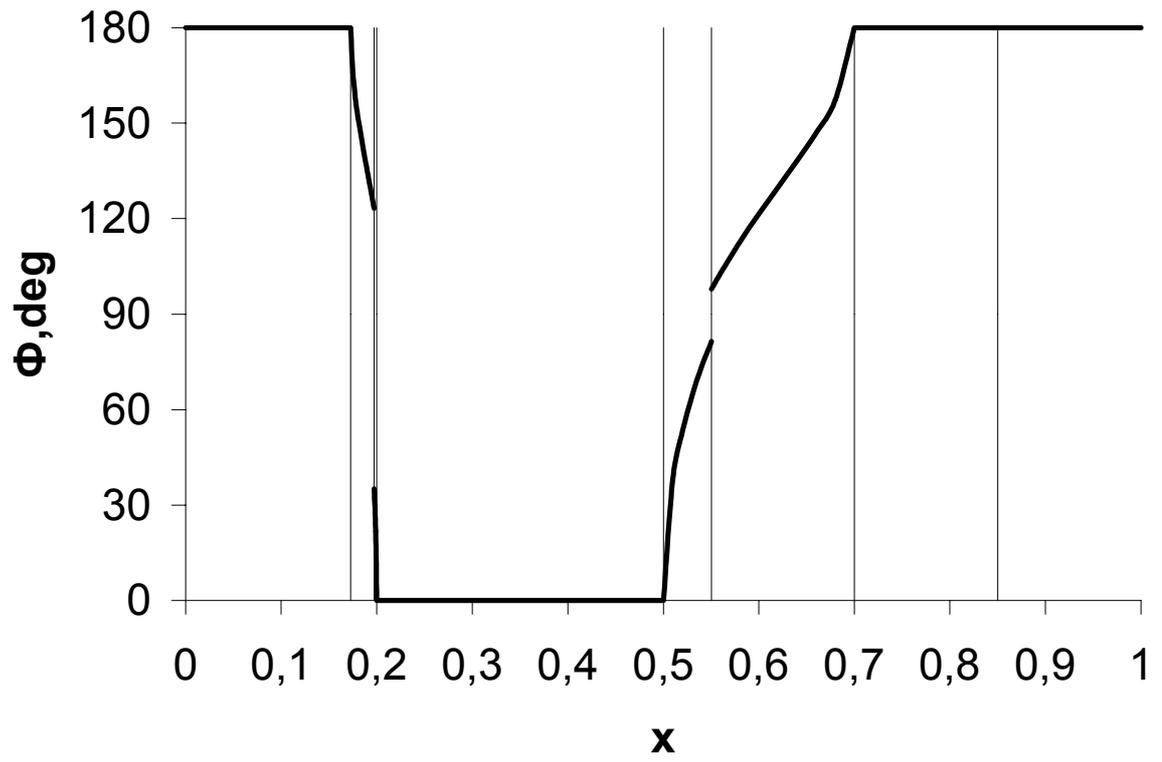

**Fig. 10**

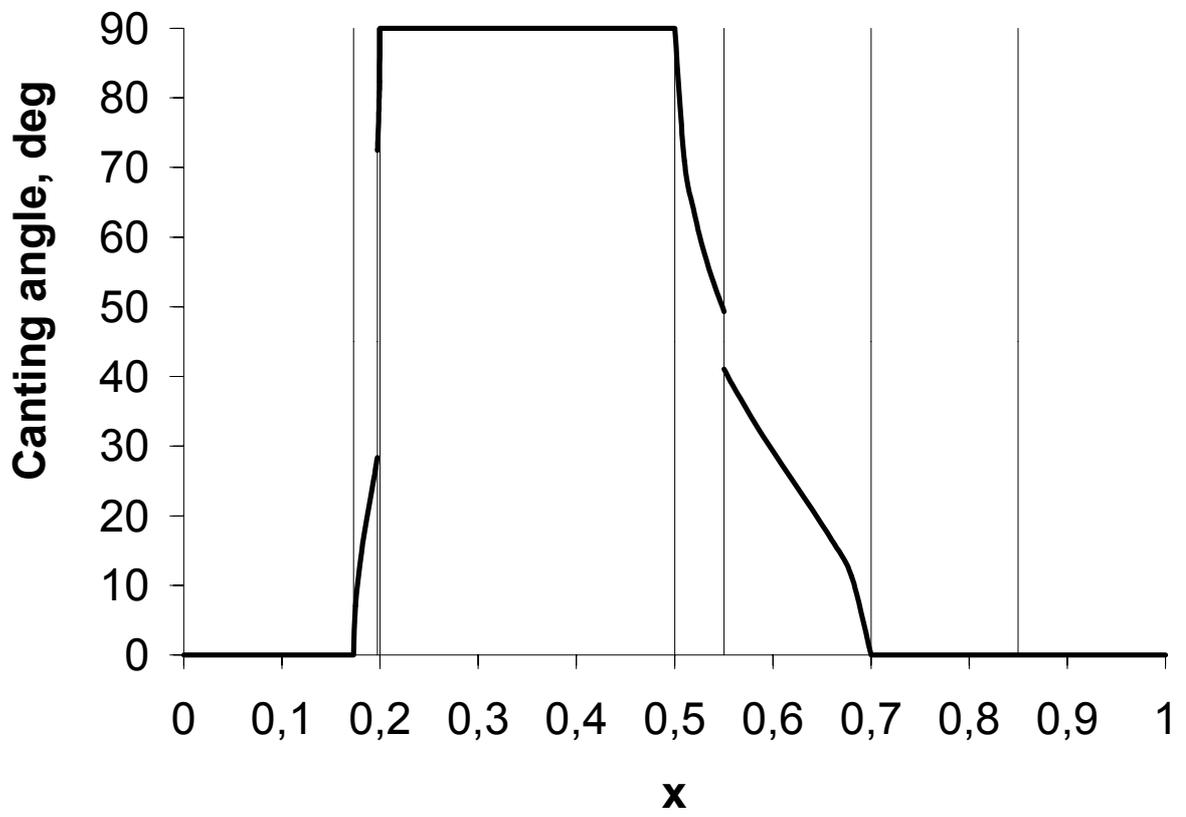

**Fig.11**

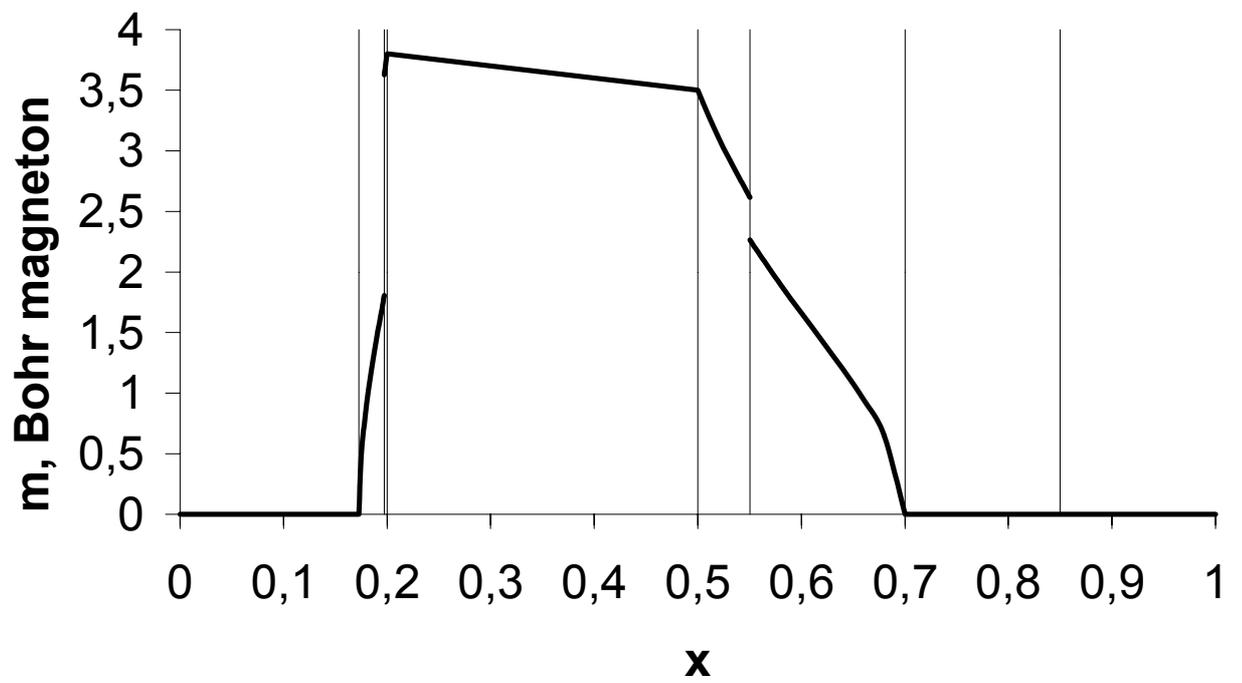

**Fig. 12**

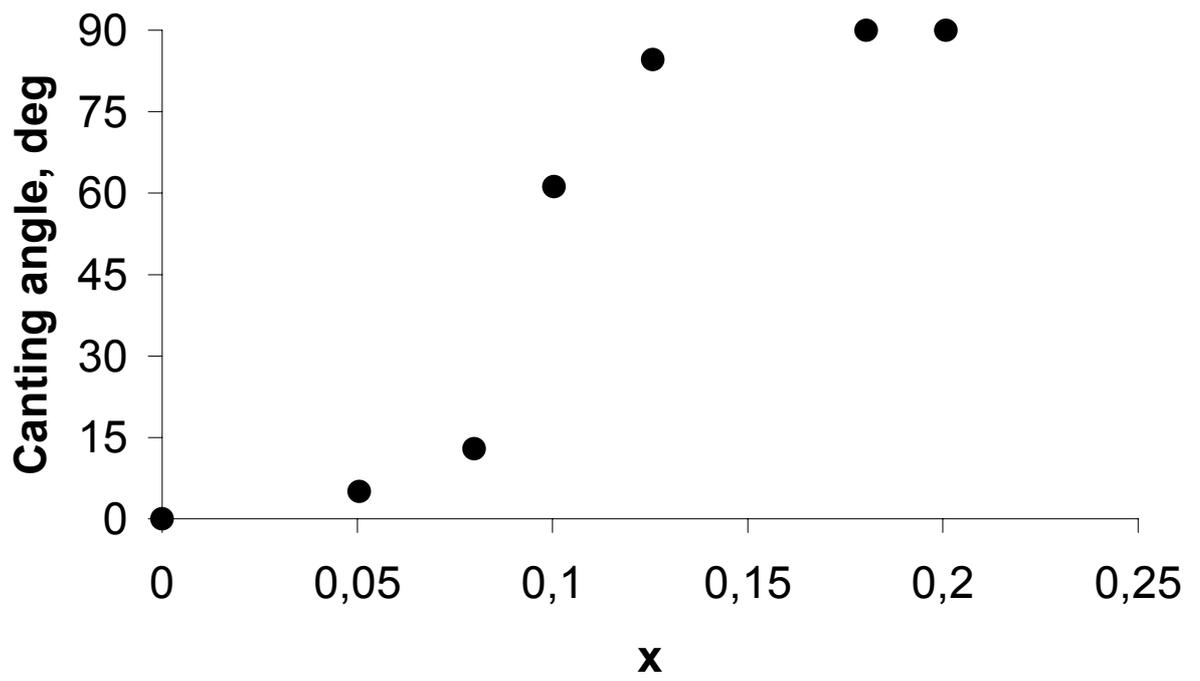

**Fig. 13**

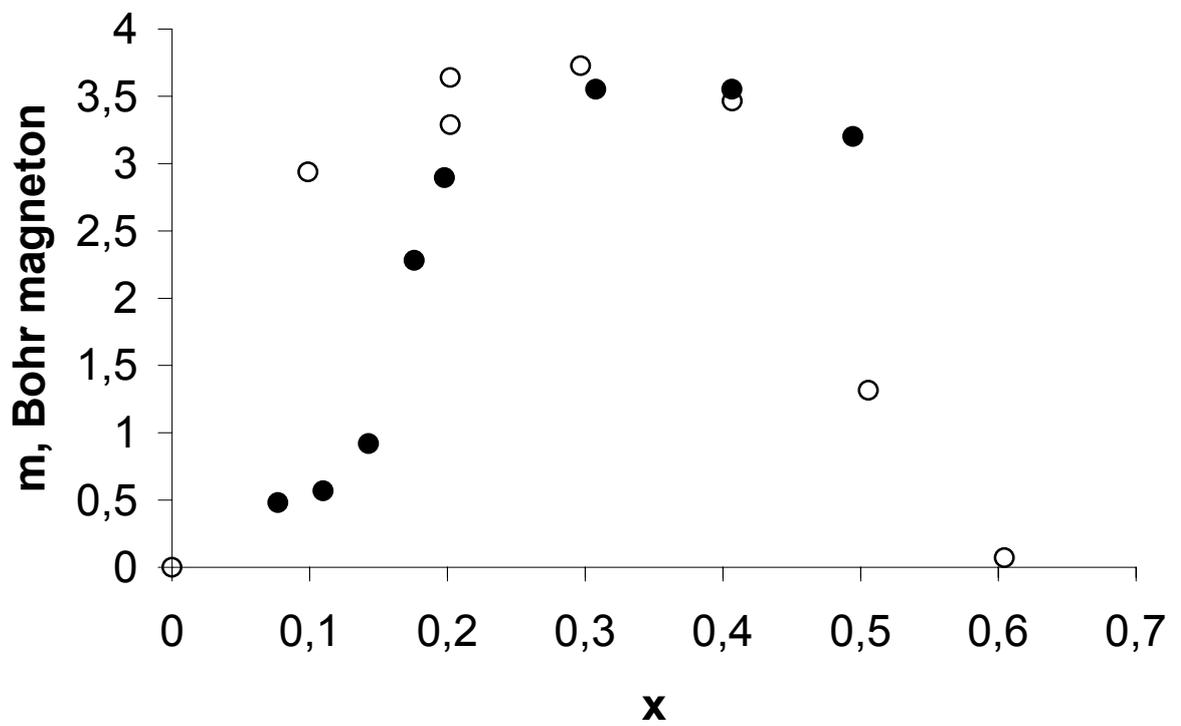

**Fig. 14**